\documentclass[aps,prd,superscriptaddress,floatfix,showpacs]{revtex4}

\usepackage{graphicx}
\usepackage{times}
\usepackage{bm}

\begin{document}

\title{The Status and Future of Color Transparency and Nuclear Filtering}
\newcommand{\orcidauthorA}{0000-0001-8181-5639} 
\newcommand{\orcidauthorB}{0000-0002-8108-8045} 
\newcommand{\orcidauthorC}{0000-0002-9296-6018} 

\author{P. Jain}
\affiliation{ Department of Space Science \& Astronomy, I.I.T., Kanpur, India, 208016;}

\author{B.~Pire}
\affiliation{ CPHT, CNRS, \'Ecole polytechnique, I.P. Paris, F91128 Palaiseau, France;}

\author{J.P. Ralston}
\affiliation{ Department of Physics \& Astronomy, University of Kansas, Lawrence, Kansas 66045;}



\begin{abstract}
40 years after its introduction, the phenomenon of color transparency remains a domain of controversial interpretations of experimental data. We review present evidence for or against its manifestation in various exclusive hard scattering reactions. The nuclear transparency experiments reveal whether short distance processes dominate a scattering amplitude at some given kinematical point. We plead for a new round of nuclear transparency measurements in a variety of experimental set-ups, including near-forward exclusive reactions related to generalized parton distribution (GPD) physics and near-backward exclusive reactions related to transition distribution amplitudes (TDA) physics.
\end{abstract}

\maketitle



\section{What Is Color Transparency?} 


Color transparency describes a universe of models and experiments where so-called ``strongly interacting'' particles can have dynamically reduced interactions which violate the rules of traditional strong interaction physics. Once upon a time, hadrons were described in a quantum mechanical
basis of asymptotic states, which interacted by the exchange of spin-0 and spin-1 mesons. The phenomenological coupling constants
of the theories were much too large to justify a perturbative description, hiding the fact the entire description was incomplete. The proposal that hadrons were {\it necessarily} complete to describe their own interactions effectively disappeared with the onset of quantum chromodynamics, QCD. And that changed everything, as history showed.

In QCD hadrons are embedded in the infinitely larger Hilbert space of a non-Abelian gauge theory, which describes how the non-Abelian charge called ``color'' flows between and inside interacting systems. Very little about the Hamiltonian or the states can be represented by the color singlet operators that create hadrons. QCD describes hadrons in terms of dramatically new configurations of quarks and gluons, some of which coherently cancel their own interactions, as we will explain. The signal is that a $QCD$ hadron can interact less than an asymptotic $S$-matrix hadron. When a
hadron interacts with a much smaller cross section than pre-QCD nuclear physics calculations would predict, we say there is evidence of ``transparency.'' In almost all cases the reactions must be exclusive, or quasi-exclusive, to develop a situation where the flow of color is narrowly channeled and then liable to destructively interfere. 

\subsection{Color Transparency Crosses Many Fields}

The field of color transparency has grown to include many experimental venues, which we briefly review here. It has grown to discover significantly different theoretical interpretations, which differ because assumptions about the constitution of hadrons are necessary to proceed. One can say that the historical origin of color transparency is the {\it Perkins effect}, published in 1955 \cite{Perkins}. Observing $e^{+}e^{-}$ pair production with emulsion targets, Perkins observed a distinct gap between the production vertex (extrapolated from tracks) to the actual onset of ionization. Perkins realized that the zero net electric charge of very localized $e^{+}e^{-}$ pairs would coherently cancel electromagnetic radiation. The pair would be invisible -- or the medium would be transparent -- until the pair had separated by a distance greater than the wavelength of radiation to be emitted to affect the emulsion. Thus Perkins invented the first ``short distance'' model of ``electromagnetic transparency''. The basic idea was re-discovered much later by Brodsky and Mueller \cite{Mueller:1982bq, Brodsky:1982kg}.  They explored the {\it proposal} that exclusive hadronic reactions with multi-GeV momentum transfer $Q$ might be dominated in $QCD$ by regions of quark wave functions that can be called ``short distance.'' Those words are loaded with theoretical interpretation, which boil down to {\it at best} quasi-localizing small color singlets on some kind of transverse scale $\Delta x_{T}$ considerably larger than optimistic dimensional analysis would suggest. The main context of short-distance dominance is associated with the model of exclusive processes originated by Farrar and Jackson \cite{Farrar:1979aw}, Efremov and Radyushkin \cite{Efremov:1979qk}, and later developed by Brodsky and LePage \cite{Lepage:1980fj}. 
The behavior of short-distance models for asymptotically large $Q$ is known, because they are designed to have a simple limit. Surprisingly, the task of showing that the large $Q$ limit of $QCD$ was the same as the limit of the models remains unfinished! 

One reason the true limits of $QCD$ is unknown is that competing models we briefly review --namely the Landshoff process, and the endpoint contribution -- do not get small compared to short-distance models in any limit. (Although, there are limits where a mechanism is introduced to make them go away.) If large $Q$ ``hard'' reactions do not select short distance, it raises the possibility that quasi-exclusive processes at large momentum transfer {\it would not} satisfy the hypothesis of being dominated by coherent cancellation of color interactions, and might not reveal any signal of color transparency. Actually that is very surprising, as if the Perkins effect would never be observed, because coherence of radiation would never happen.

Just as in Perkins' emulsion experiments, the {\it medium} used to measure color transparency effects cannot be separated from the phenomenon. Perkins had the first observation of the {\it survival probability}, which in his experiment was a coherence-determined region of particles surviving before dissipative interactions could begin. By involving coherence and entanglement, the survival probabilities of color transparency experiments are quintessential testing grounds of quantum mechanics. {\it Nuclear filtering} \cite{Jain:1995dd} is the topic that recognizes a quantum mechanical measurement of 
a process affects the process. In particular, when particles survive without much interaction, a process has selected them with preference for less interaction. Nuclear filtering is a quantum measurement feature that is intimately related to color transparency, while it also has a classical analog. Let us discuss this briefly. 

\subsection{Survival of the Smallest} 

One way to think about color transparency in nuclear targets begins with a distribution of interacting systems in an absorbing medium. The notion of a hypothetical distribution is actually not easily separated from a distribution of entities produced by an interaction. In any event, consider classical particles of variable radius $a$ interacting with a medium with absorption cross section $\sigma(a) = \pi a^{2}$. Let $n$ be the number density of targets, and $L = 1/n\sigma(a)$ be the mean free path. Given a particular size $a$, the probability to cross a distance $z$ is proportional to $exp(-z/L)= exp(-\pi z a^{2}n)$. The exponential dependence is familiar. But in the present context, it is a conditional probability $P(z \, | \, a)$, which must be combined with the probability to find radius $a$, denoted $P(a)$. To make a computable model, suppose $P(a)$ is a Gaussian times a power, $P(a) \sim a^{m}exp(-a^{2}/2 a_{0}^{2})$. The peak of this distribution will be at a computable point $a \sim a_{*}$. Then compute 
\begin{equation}
 P(z) =\int da \, P(z \, | \, a)P(a)  = \int_{0}^{\infty} da \,  exp(-\pi z a^{2}n) a^{m}exp(-a^{2}/2 a_{0}^{2}) \sim {1\over ( 1+ 2 \pi  a_{0}^{2}nz)^{(m+1)/2}}. \nonumber 
 \end{equation} 
This is quite an interesting result. Instead of decreasing exponentially with depth, the survival probability decreases like an {\it inverse power}. That should produce an impressive signal of the distribution of interacting systems, compared to a single interaction cross section distributed by $\delta(a -a_{*} )$. Yet to detect this experimentally, the medium must be thick enough to distinguish the exponential fall from the power law decrease. This is known but not always recognized. In the context of reactions involving quasi-elastic collisions with nuclear targets, there is a strong trend for a small nucleon number $A$ to reduce experimental discriminatory power, and for large $A$ to increase it. This contradicts an occasional perception that smaller $A$ nuclei are understood better than larger $A$ and should be better experimental topics. As a result some inconclusive results have been found for rather small $2 \lesssim A \lesssim 12 $. But much of what is understood at small $A$ is in the hadronic basis of proton and neutron wave functions, not the basis of quarks and gluons where color transparency originates. Many experiments involving $A>>1$ are still waiting to be done. When one is exploring new territory, why not measure everything that can be measured?

\section{Highlights of Two Experiments} 

\begin{figure}[ht]
\begin{center}
\includegraphics[width=4in]{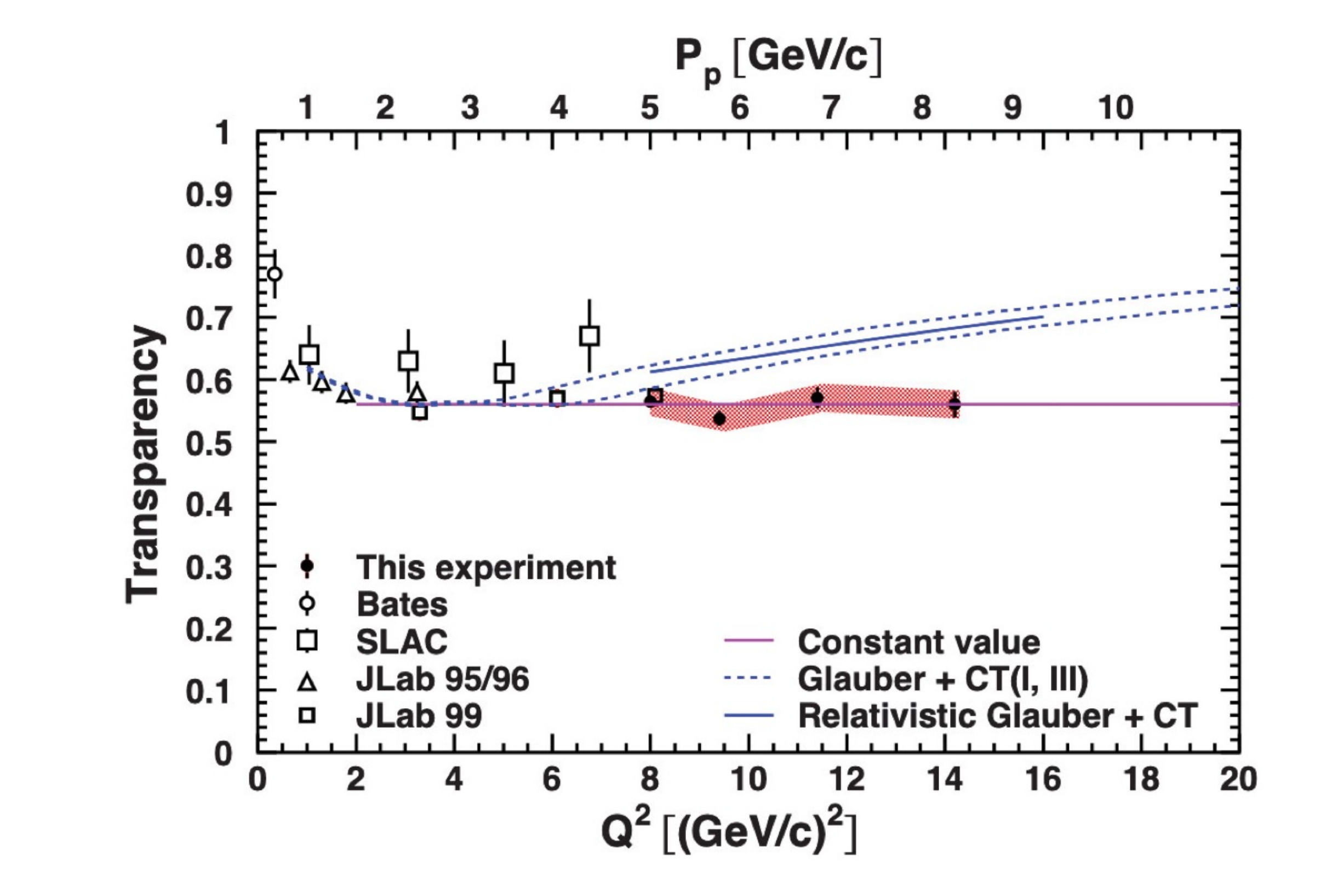}
\caption{ \small The $Q^{2}$ dependence of the transparency ratio for $^{12}$C measured by Bhetuwal et. al. and previous JLAB experiments \cite{Makins:1994mm,ONeill:1994znv,Abbott:1997bc,Garrow:2001di,Bhetuwal:2021}.}
\label{fig:Bhetuwal}
\end{center}
\end{figure}

Two experimental programs in color transparency stand out as being most informative, and possibly most contradictory. A series of quasi-exclusive A$( e, \, e'p)$ experiments at Jefferson Lab \cite{Makins:1994mm,ONeill:1994znv,Abbott:1997bc,Garrow:2001di,Bhetuwal:2021}
for a $^{12}C$ nucleus was recently extended to the range $ 8\lesssim Q^{2}\lesssim 14.2$ GeV$^{2}$. The ``transparency ratio'', (here defined by the measured rate compared to a partial wave impulse approximation (PWIA) calculation with Glauber interactions) was determined to be practically constant over the range of the experiment. See Figure \ref{fig:Bhetuwal}. The absence of any signal of color transparency has led to an apparent consensus that a short-distance model describing the process has been ruled out. While a short distance model is often identified as being the same as ``$QCD$,'' other models in $QCD$ have no short-distance feature, including in particular the Feynman or endpoint mechanism discussed below. 

\begin{figure}[h!]
\begin{center}
\includegraphics[width=4in]{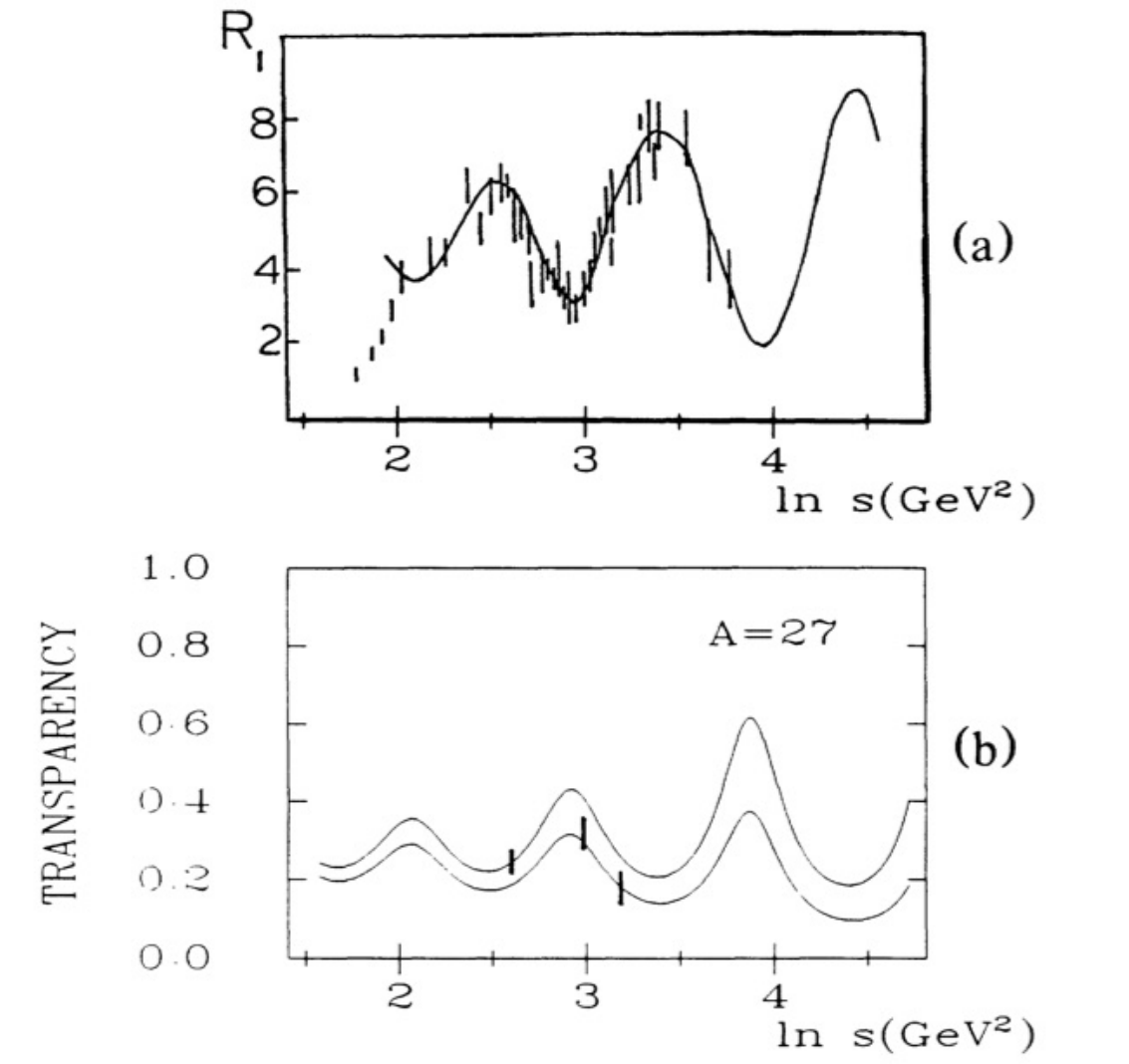}
\caption{ \small The $s$ dependence of the cross section $s^{10}d\sigma/dt$ for $pp \rightarrow p'p''$ in free space (top) compared to the BNL transparency ratio $T(s)$, which uses it as the denominator. Multiplying $T(s)$ by the denominator finds constant $s$ dependence, which is consistent with nuclear filtering removing an interfering amplitude with an $s$-dependent relative phase. Figure from Ref. \cite{Ralston:1988rb} with permission.}
\label{fig:wiggles}
\end{center}
\end{figure}

Another highly informative experiment is BNL E834, measuring the center of mass energy dependence of fixed angle quasi-elastic knockout of protons from nuclei, $pA \rightarrow p'p''(A-1)$. The quasi-elastic kinematics veto events where pions are produced. The first announcement of the experiment caused great excitement due to the appearance of a ``bump'' near 10 GeV in the beam energy dependence of the transparency ratio for $A$=12, 27, 64, and 204. The decrease of this ratio above 10 GeV ruled out classical expansion models \cite{Farrar:1988me}. This may not have been appreciated, at first, while it was settled decisively with the larger data set of the experiment's final report \cite{Aclander:2004zm}. The famous ``bump'' has been explained by two facts: First, the BNL transparency ratio $T(s)=d\sigma/dt(pA \rightarrow p'p''(A-1))/\left( Z d\sigma/dt(pp \rightarrow p'p'' ) \right)$. The denominator is experimental data, and not a calculation as defined by JLAB physicists. Here $s$ is the Mandelstam center of mass squared energy on a proton target. Second, the energy dependence of denominator has oscillations 180$^{o}$ out of phase with $T$. This is shown in Figure \ref{fig:wiggles} from Ref. \cite{Ralston:1988rb,Jain:1995dd}. Multiplying $T(s)$ by the denominator shows that $s^{10}d\sigma/dt(pA \rightarrow p'p''(A-1))$ is constant with $s$. It is hard to escape the conclusion that $pp \rightarrow p'p''$ in free space oscillates due to interference of different amplitudes with an $s$-dependent relative phase. That interference then disappeared in the BNL experiment by nuclear filtering. In fact, the free space oscillations had previously been identified \cite{Pire:1982iv} with interference of a short-distance component, and a computable chromo-Coulomb phase \cite{Ralston:1982pa} of the Landshoff process \cite{Donnachie:1979yu}, to be defined in a moment. It is beautifully consistent that the Landshoff process is not a short distance one, but has well separated quarks that nuclear filtering would naturally attenuate. Since those days {\it endpoint model} processes (see below) which are equally susceptible to nuclear filtering have emerged as viable candidates.

While the BNL experiment has the natural explanation just given, it is not without controversy, and in fact rests on rather old $pp$ fixed angle scattering data that has never been repeated. To this day the issues are alive with controversy, and partisan debate. The importance of those experiments, and the rather fundamental character of the QCD chromo-Coulomb phase shift believed to cause the oscillations, represent important opportunities that JPARC's program should explore.


\section{Models of Short and Not-So-Short Distance Processes} 
\subsection{Momentum vs coordinate space descriptions}
\begin{figure}[ht]
\begin{center}
\includegraphics[width=5in]{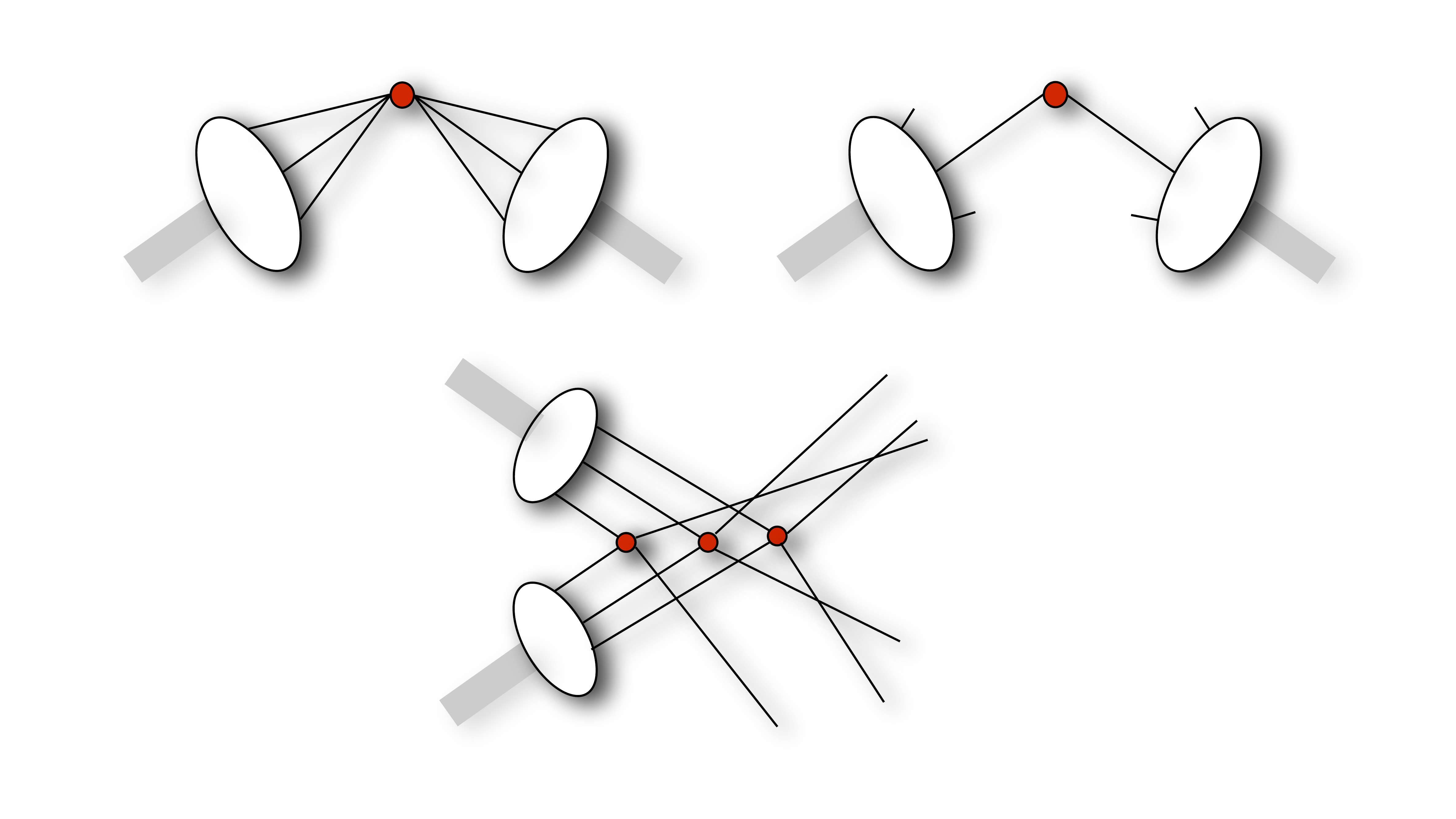}
\caption{ \small Real space pictures associated with different momentum-space integration regions of models of exclusive processes. {\it Top Left:} A short distance model effectively has participating quarks meeting at a hard scattering (red dot) with points separated spatially by order $1/Q$. {\it Top Right:} An endpoint model has a single hard scattering. Low momentum quark spectators transfer from one struck hadron to another. Since finding soft, non-participating quarks is rare in proportion to how many are soft, this process happens to have the same ``quark counting'' rules as a short distance model. {\it Bottom}: A Landshoff process (bottom) has quarks acting completely independently, and flying off at random. The coincidence of final state quarks accidentally traveling parallel to form a final state hadron is calculated from a phase space integral. }
\label{fig:ThreeProcesses}
\end{center}
\end{figure}

The great complication of $QCD$ is the existence of many {\it infinities} of dynamical degrees of freedom. While we might prefer to simply forget them, a valid theoretical calculation needs to explain what has been done with them. Here we give readers a guide to how calculations are translated into the words describing them. A guide is needed because calculations are done in the momentum space or participating quarks, while the words refer to coordinate space. In a nutshell: 
\begin{itemize} 
\item Perturbative $QCD$ describes interactions of quarks and gluons at ``large'' momentum transfers $Q^{2} \gtrsim$ GeV$^{2}.$ \item $pQCD$ descriptions of hadron scattering always involve unknown elements, which are the wave functions of quarks and gluons in the hadrons. These wave functions are field-theoretical entities that cannot be developed from a non-relativistic few quark picture. 
\item All models of large momentum transfer involve a strictly limited number of participating quarks. The lowest order building block is $2\rightarrow 2$ quark-quark (or quark-antiquark) scattering via one-gluon exchange. Models differ in how the building blocks are put together. 
\item Amplitudes are calculated by integrating over internal momenta. When models are verbally described as ``involving this or that process,'' it is really a translation of an approximation concentrating on a momentum region that can be identified with the process. A short-distance model is one that actually asserts certain regions of high momentum transfer dominate the calculation. A not-so-short distance processes may integrate over regions of high momentum transfer, along with other regions whose scale is set by the wave function of the hadron. 
\item With this information, models relevant to color transparency can be understood with the real-space cartoons of Figure \ref{fig:ThreeProcesses}. In the top panel, a short distance model (left) has emphasis on participating quarks meeting at a hard scattering (red dot) with points separated spatially by order $1/Q$. An endpoint model (right) has a single hard scattering, with soft, long wavelength quarks not participating, and merging from one struck hadron to another. This process happens to have the same ``quark counting'' rules as a short distance process, because finding soft, non-participating quarks is rare in proportion to how many are soft \cite{Dagaonkar:2014yea}. A Landshoff process (bottom) has quarks acting completely independently, and flying off at random. The coincidence of final state quarks accidentally traveling in parallel to form a final state hadron is calculated from the phase space integral. Perhaps surprisingly, this process dominates the processes $\pi \pi \rightarrow \pi \pi, \, \pi p  \rightarrow \pi p ,p  p  \rightarrow p  p$,  even when radiative (Sudakov) factors are applied
to account for the lack of final state radiation\cite{Mueller:1981sg, Botts:1989kf, Li:1992nu}.
\end{itemize}

\subsection{More About the Endpoint Process} 

The endpoint model was in fact the first model proposed for exclusive processes, originally by Feynman in 1969 \cite{Feynman69}. As explained above, this involves only one hard scattering. Essentially one of the quarks carries most of the hadron momentum with the corresponding longitudinal momentum fraction $x\rightarrow 1$. This quark undergoes a hard scattering and the resulting amplitude has a direct dependence on the nature of the hadronic wave function in the limit $x\rightarrow 1$. Given this dependence, it is not immediately apparent how such a process might lead to the observed scaling laws at large $Q^2$. It was shown in \cite{Dagaonkar:2014yea} that the observed scalings laws follow if we assume a power dependence on the momentum fraction variables $x_i$. For example, for  pion we obtain $F_\pi\sim 1/Q^2$ if we assume that the wave function $\Phi(x,\vec k_T)\sim x(1-x)\sim (1-x)$ for $(1-x)$ of order $\Lambda_{QCD}/Q$. For proton, we assume that quark 1 carries most of the momentum, i.e. $x_1\rightarrow 1$, while $x_2$ and $x_3$ are of order $\Lambda_{QCD}/Q$. We obtain the observed dependence $F_1\sim 1/Q^4$, if the wave function goes as $x_2 x_3$ in the limit $x_1\rightarrow 1$. Given such an $x$ dependence of the wave function, the end point model also remarkably leads to the power laws observed in hadron-hadron scattering for fixed angle scattering assuming $s\sim |t|>> 1$ GeV$^2$ . In this case a leading quark from each hadron undergo hard scattering with one another with exchange of hard gluons.  The remaining quarks again merge into the final state hadrons as in the case of form factor. It is quite interesting that this mechanism naturally generalizes to another regime, i.e. $s>>|t|>> $ GeV$^2$ at fixed $s$. In this case, for $pp\rightarrow pp$ the end point model predicts $d\sigma/dt\sim 1/|t|^8$, in excellent agreement with data \cite{Donnachie:1979yu}. 

The endpoint model relates the $x$ dependence of the wave function in the limit $x\rightarrow 1$ to the observed scaling laws in exclusive processes. Hence it provides an opportunity to experimentally deduce the hadron wave function in this limit. Being a soft
mechanism, it does not, for example, predict the hadron helicity conservation rule. This rule is predicted by the hard scattering mechanism \cite{PhysRevD.24.2848}, but not observed in data. It also naturally leads to the prediction, $F_2/F_1\propto 1/Q$ \cite{Dagaonkar:2015laa}, in agreement with data, while the short distance model would predict $F_2/F_1\propto 1/Q^2$. Furthermore, it does not predict the phenomenon of color transparency, at least in its simplest form, and hence is nicely consistent with experimental results \cite{Makins:1994mm,ONeill:1994znv,Abbott:1997bc,Garrow:2001di,Bhetuwal:2021}. It is not, however, clear how the endpoint model may explain the $pA$ data on color transparency \cite{Carroll:1988rp}. Perhaps in this case nuclear filtering plays an important role and the short distance contributions may not be negligible in nuclear medium \cite{Jain93,Jain:1995dd}. However the phenomenon needs further study in order to assess the relative importance of different contributions.

\section{The GPD domain : near forward exclusive electroproduction and related processes}

Near-forward exclusive photon or meson electroproduction processes have been the subject of intense theoretical and experimental studies \cite{Diehl:2003, Kumericki:2016ehc}. Most of the available data are now interpreted in terms of a collinear QCD factorized amplitude, where generalized parton distributions (GPDs) are the relevant hadronic matrix elements. The topical reactions - deeply virtual Compton scattering (DVCS) and its timelike extension timelike Compton scattering (TCS) - where quite detailed  phenomenological studies have been conducted, have concluded that available data are compatible with early scaling, and leading twist  factorization. These processes, which are purely electromagnetic at Born order, are not directly relevant to color transparency studies, since the outgoing nucleon is not attached to the hard amplitude but rather to the fully non-perturbative hadronic matrix element (the GPDs). This fact does not constrain the nucleon to be in a small size configuration when it scatters.

Meson electroproduction 
\begin{equation}
    \gamma^* N \to M N' \,,
\end{equation}
has been shown to obey the same factorization properties as DVCS and TCS at large $Q^2$. This opens the possibility to perform  study of nuclear transparency for the outgoing meson (but still not for the outgoing nucleon) which is attached, through its distribution amplitude (DA) to the hard amplitude.  Ref. \cite{Clasie:2007aa,CLAS:2012tlh} indeed revealed a growth of the nuclear transparency ratio indicative of an early on-set of the scaling regime \cite{Cosyn:2007er} for both $\pi$ and $\rho$ electroproduction. In the $\pi$ case, this may however look contradictory to the non-dominance of the leading twist pion production amplitude revealed by the small value  of the polarization ratio $\sigma_L / \sigma_T$ for this reaction \cite{JeffersonLabHallA:2016wye, JeffersonLabHallA:2020dhq}. 

The presence of leading twist and higher twist contributions to the $\pi$ meson electroproduction amplitude opens the way to a nuclear filtering interpretation. To check such interpretation, a polarization sensitive experiment needs to be done in the nuclear target case as it was performed in the proton target case. Nuclear filtering then predicts a growth of the $\sigma_L / \sigma_T$ ratio when the size of the nucleus increases as a consequence of the {\it survival of the smallest} principle. Let us however point out that some higher twist contributions to this process may also imply a small size meson configuration. 

Since the late scaling of the $\pi$  meson electroproduction may be related to the peculiar chiral nature of the $\pi$ meson, it is mostly important to prepare CT experiments for other mesons ($\eta, K, ...$) and at higher energies.

Similar reactions may be performed with meson beams at JPARC and COMPASS at Cern. The amplitudes for exclusive Drell-Yan processes such as
\begin{equation}
    \pi^-(p_\pi) p(p_N) \to \gamma^*(q) n(p_n) \,,
\end{equation}
indeed factorize at large $(Q^2 =q^2)$ and small $t = (p_N -p_n)^2$ in a hard part convoluted to GPDs \cite{Berger:2001zn}. Such studies have been shown to be feasible at JPARC \cite{Sawada:2016mao}. In this case, the initial $\pi^-$ is attached to the hard scattering through its DA and should thus be subject to CT effects, now understood as the decrease of initial (rather than final) state interactions.

\section{Color transparency for backward scattering processes}
\subsection{Backward electroproduction}
It is quite natural to extend the study of exclusive electroproduction processes in the complementary near backward region, where the kinematics of the process
\begin{equation}
 \gamma^*(q) N(p_N) \to M(p_M) N'(p_{N'}) \,,   
\end{equation}
is restricted to the region where $- u = -(p_M - p_N)^2 \ll Q^2 = -q^2$ is near its minimal value~\cite{Gayoso:2021rzj}. One thus  describes the amplitude of this process at large $Q^2$ in a collinear QCD factorization scheme \cite{Frankfurt:1999fp, Pire:2004ie, Pire:2005ax}, where nucleon to meson transition distribution amplitudes (TDAs) replace the GPDs as the relevant hadronic matrix elements \cite{Pire:2021hbl}. Indeed, the first JLab experimental studies \cite{Park:2017irz, Li:2019xyp, Diehl:2020uja} of this new domain at rather moderate values of $Q^2$ point toward an early onset of the scaling regime.

Looking for color transparency effects in such kinematics opens the possibility to study the fast produced nucleon propagation in the target nucleus. Indeed, in these backward regime, the hadron attached to the hard scattering is the produced nucleon, while the  "backward" meson is attached to the non-perturbative hadronic matrix element and as such is not constrained to be in a small size configuration, and thus  likely to be subject to full final state strong interactions. A typical example is backward electroproduction of a $\pi^0$ meson, as detailed in Ref. \cite{Huber:2022wns}.
\subsection{Electromagnetic processes at PANDA}
\begin{figure}
\centering
\includegraphics[width=0.9\textwidth]{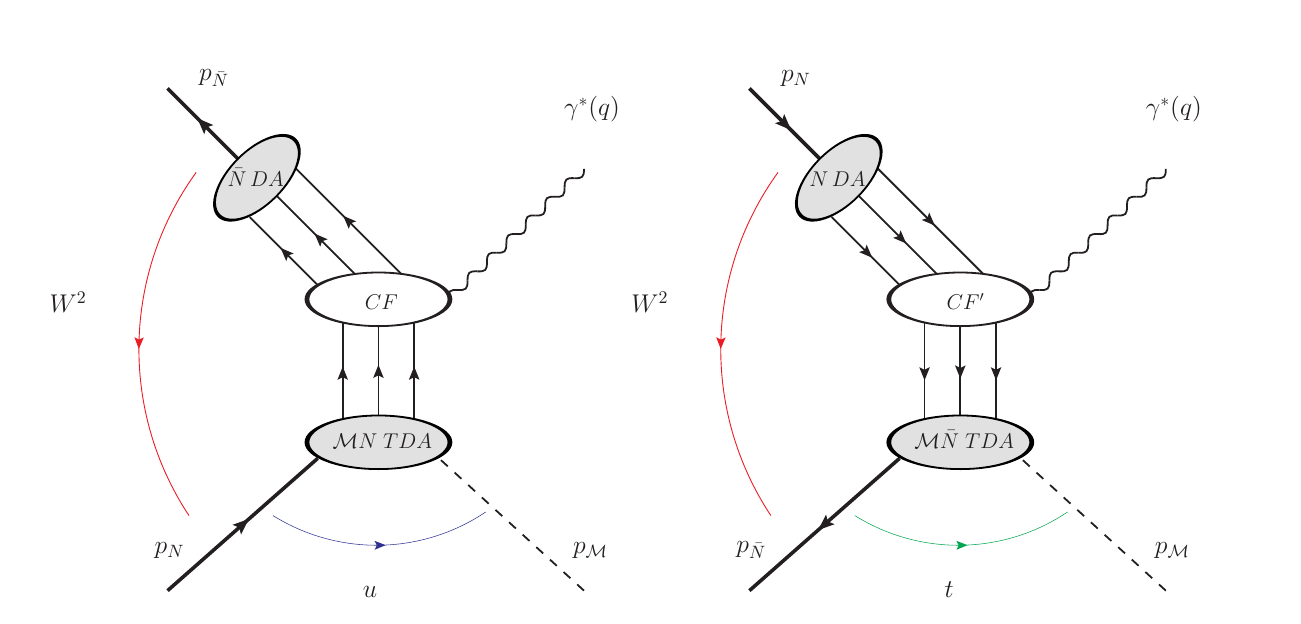}
\caption{\label{fig:PANDA} The annihilation process  $\bar N N \to \gamma^* M$ at PANDA is sensitive to the TDA factorization dynamics in both near-backward (left panel) and near-forward (right panel) kinematics.}
\end{figure}

The antiproton beam experiments to be carried by the PANDA detector \cite{PANDA:2009yku} at FAIR allow to test the TDA factorization framework \cite{Pire:2004ie, Lansberg:2012ha} in crossed exclusive processes such as
\begin{equation}
    \bar P(p_{\bar p}) N(p_N) \to \gamma^*(q) M (p_M)
\end{equation}
at large timelike $Q^2=q^2$ and small $u=(p_M- p_{N})^2$ (backward meson) or small $t=(p_M- p_{\bar p})^2$ (forward meson) kinematics, as depicted on Fig.\ref{fig:PANDA}. In both cases, the (anti)nucleon is attached to the hard part through its distribution amplitude, thus restricting its transverse extension to small $O(1/Q)$ size. There is however a big difference in the relative velocity of this state with respect to the nucleus, and one thus  anticipate a much stronger nuclear transparency effect in the near backward kinematics than in the complementary one. Such an experiment has thus the unique capability to disentangle small size configuration production effects from  transverse expansion consequences.
\subsection{CT in  backward  processes with meson beams}

 The  backward exclusive Drell-Yan reaction 
\begin{equation}
    \pi(p_\pi) N(p_N) \to \gamma^*(q) N' (p'_N) \to l^+ l^- N' (p'_N)
\end{equation}
at large timelike $Q^2=q^2$ and small $u=(p_\pi- p'_{N})^2$ as well as charmonium production
\begin{equation}
    \pi(p_\pi) N(p_N) \to J/\psi(q) N' (p'_N)
\end{equation}
in the same kinematical region, can also be described \cite{Pire:2016gut} in a collinear factorization framework with $\pi \to N$ TDAs as the relevant hadronic matrix elements. Here the initial nucleon supplies the necessary small size hadronic configuration and color transparency should manifest itself through the decrease of initial state interactions. A feasibility study needs to be done for both processes with the expected luminosity of the high energy secondary meson beam at JPARC. 

Before ending this section, we point out that there exist various Regge interpretations of electroproduction data  \cite{Yu:2018ydp, Kaskulov:2010kf, Laget:2021qwq}, which essentially deny any color transparency effect. Furthermore, to the best of our knowledge, neither the forward nor the backward scattering processes have been studied within the framework of the endpoint model. Such a study would be very useful in order to firmly establish the theoretical predictions for color transparency in these processes. 

\section{Conclusion}
Color transparency is a fascinating subject at the crossroad of perturbative QCD and nuclear physics. Present nuclear transparency measurements show that the concept can be probed in different set-ups. Much experimental work is still needed to enlighten its possibility to signal the dominance of small distance processes at a given kinematical point. We proposed here various processes to be scrutinized in the near future  at existing or projected facilities. The 2021 workshop on  {\it the Future of Color Transparency and Hadronization studies at Jefferson Lab and beyond} indeed opened new ways to this exploration.

\section*{Acknowledgements}
We acknowledge numerous fruitful discussions with W. Cosyn, K. Semenov-Tian-Shansky and L. Szymanowski.  

\bibliographystyle{alpha}
\bibliography{sample}

\end{document}